\documentclass[review,5p,twocolumn]{elsarticle}
\usepackage{color,amsmath}

\usepackage{lineno,hyperref}

\modulolinenumbers[0]

\journal{Journal of \LaTeX\ Templates}

\begin{document}

\begin{frontmatter}

\title{Hybrid atom-photon entangling gates via Gaussian soft control}

\author{Wanrang Yu}\author{Qiuyu Yin}\author{Yanzhao Liang}\author{Ning Ji}\author{Thibault Vogt\corref{cor1}}\cortext[cor1]{Corresponding author}
\ead{ttvogt@mail.sysu.edu.cn}
\address{School of Physics and Astronomy, Sun Yat-Sen University, Zhuhai 519000, China}
\begin{abstract}
Hybrid atom-photon gates play an important role for the realization of a quantum interface capable of mapping atomic states to photons for communication across quantum networks. Here, we propose a feasible theoretical scheme for implementing a hybrid atom-photon controlled-Z gate between an atom and a microwave photon in a superconducting coplanar waveguide resonator based on the Gaussian soft control technique. 
The gate protocol employs a classical auxiliary field that induces an atomic transition between one state of the atomic qubit and Rydberg states for obtaining strong coupling of the atom and microwave resonator.
By tailoring the amplitude of this field with Gaussian temporal modulation, the gate performances are improved in various aspects. Numerical simulations demonstrate that the controlled-Z gate based on Gaussian soft control is resilient to the variation of the atom-photon coupling strength, deviation in the gate time, and less sensitive to the Rydberg level shifts caused by stray electric fields.
\end{abstract}


\end{frontmatter}


\section{Introduction}

Quantum networks are essential for long-distance quantum communication~\cite{PSAB1012,SZL367}, distributed quantum computation~\cite{JTS062323,PHB259}, and quantum metrology~\cite{GE0436,GLM222}. Some experiments have separately demonstrated the capabilities of executing fast high-fidelity quantum-logic gates and efficient distribution of quantum information among remote quantum nodes~\cite{KHJ1030,WED9288}, but achieving them simultaneously in a single physical system is still challenging. Quantum hybrid systems can serve as new platforms where different types of physical systems are combined together to leverage the unique strengths of each of them for implementing novel functionalities~\cite{KBP3866,CLB257}.
 For example, superconducting (SC) circuits operating at microwave frequencies are characterized by their fast gate time and scalability, but their decoherence time is rather short~\cite{BGG025005,BPMN094,BASM256}. On the other hand, atomic systems can offer very long coherence times~\cite{DMR339}. A particularly promising hybrid quantum system is integrating SC quantum circuits with neutral atoms to achieve the capabilities of both fast processing and long coherence time. The transitions between Rydberg energy levels exhibit high sensitivity to microwave fields in the frequency range corresponding to that employed in circuit quantum electrodynamics~\cite{ACSP012,SWTG2313}. This characteristic makes Rydberg states highly suitable for mediating the coupling of a superconducting coplanar waveguide (SCW) resonator with atoms on an integrated atom chip~\cite{YDAV062,VZKH043}. Recent developments in chip-based experiments have demonstrated the coherent manipulation of Rydberg states close to chip surfaces~\cite{KMGL013,GAMJ0434}.\\
\indent For the SCW resonator-atom hybrid quantum system, hybrid quantum gates have the ability to interface different types of qubits, providing an essential foundation for the realization of more complex quantum network structures~\cite{LYLMY013,PAB010,SXF042,LKYZT296}.  
Schemes using constant-amplitude field pulses are widely employed for implementing fast and robust quantum gates, but they usually require multistep operations and are sensitive to control errors~\cite{WJLW012}. Thus, there has been a significant focus on developing pulse engineering techniques~\cite{GGTV133,ZCHS042,KTSM53,WHSA06,LYSP034} to accomplish quantum tasks with high fidelity and robustness. The Gaussian soft control (GSC) technique  introduced by Hasse et al.~\cite{HJFWC050} is a typical representative for pulse engineering. It provides an effective tool to suppress non-resonant interactions within a target system. This technique can be used for state preparation~\cite{HJMW032}, construction of quantum gates~\cite{YHDXQ2541,ZQLW095,YMRCS580}, and quantum sensing~\cite{WZYLJE2000}. \\
\indent In this article, we study a hybrid atom-photon controlled-Z (CZ) gate based on the GSC technique in a hybrid SCW resonator-atom quantum system. The proposed scheme requires only one auxiliary driving field to induce the CZ gate. 
For a rectangular pulse of the driving field, specific ratios $g/\Omega$ of the atom-photon coupling strength $g$ and Rabi frequency of the driving field $\Omega$ permit the realization of the atom-photon CZ gate~\cite{SXF042}. However, using this constant-amplitude pulse shows weak resilience to parameter errors.
To address these limitations, we change the time-independent rectangular driving field into a time-dependent one with Gaussian temporal modulation to maintain efficient resonant couplings among target levels and suppress unwanted non-resonant contributions. Compared with the case of a constant amplitude, the gate scheme based on GSC shows more robustness against parameter errors. In particular, it features high resilience against the effect of stray electric fields, an inherent problem of SCW resonator-atom hybrid systems~\cite{KMGL013,BH}. Our work is promising for future quantum network architectures involving different quantum systems.

\section{Model and construction of the CZ gate}
The physical model utilized for implementing the hybrid atom-photon quantum gate is illustrated in Fig.~1(a), where an atom is confined near the surface of a SCW resonator. The SCW microwave field and the atomic energy levels required for realizing the hybrid atom-photon gate are depicted in Fig.~1(b). 
\begin{figure}
	\centering\includegraphics[width=\columnwidth]{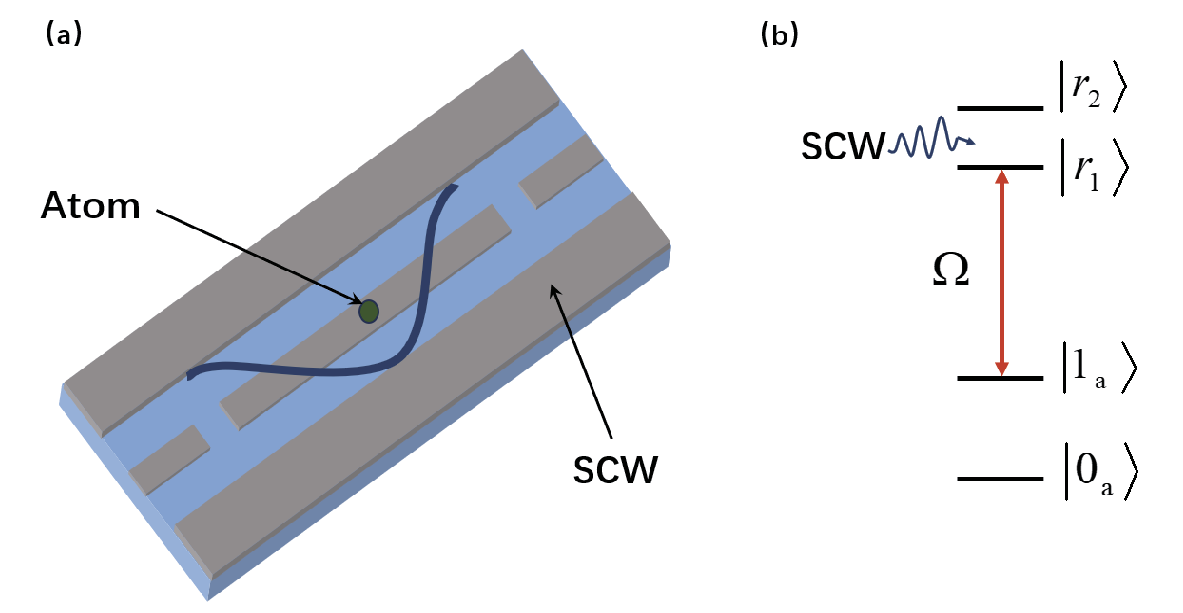}
	\caption{(a)~A diagram of the experimental setup for the hybrid quantum gate scheme showing an atom interacting with a superconducting coplanar waveguide (SCW) resonator. (b)~The energy-level configuration of the atom used for realizing the CZ gate.}
	\label{fig1}
\end{figure}
The atom possesses two ground states  $|0_a\rangle$ and $|1_a\rangle$, along with a pair of Rydberg states $|r_1\rangle$ and $ |r_2\rangle$. A classical laser field with Rabi frequency $\Omega$ drives the transition $|1_a\rangle\leftrightarrow|r_1\rangle$. Given a photon in the SCW resonator, the microwave transition $|r_1\rangle\leftrightarrow|r_2\rangle$ is resonantly coupled to the quantized cavity field with coupling strength $g$. In the rotating frame and under the rotating wave approximation (RWA), the interaction Hamiltonian of the hybrid system can be expressed as ($\hbar=1$)
\begin{eqnarray}\label{e1}
	H_I&=&H_g+H_{\Omega},\nonumber\\
	H_g&=&g(|r_2\rangle\langle{r_1}|a+|r_1\rangle\langle{r_2}|a^{\dagger}),\nonumber\\
	H_{\Omega}&=&\frac{\Omega}{2}(|1_a\rangle\langle{r_1}|+|r_1\rangle\langle{1_a}),
\end{eqnarray}
where $a$ ($a^{\dagger}$) is the annihilation (creation) operator of a photon in the SCW resonator. We use $|0_m\rangle$ and $|1_m\rangle$ to label the photon number in the SCW resonator.

If the atom is initially in state $|0_a\rangle$, states $|0_m0_a\rangle$ and $|1_m0_a\rangle$ are not altered since they are not coupled to the driving auxiliary field, as shown in Figs.~2(a) and 2(b).
If the system is initially in state $|1_m1_a\rangle$, it evolves in the finite subspace \{$|\phi_1\rangle=|1_m1_a\rangle$, $|\phi_2\rangle=|1_mr_1\rangle$, $|\phi_3\rangle=|0_mr_2\rangle$\}. The couplings between these three states, $|1_m1_a\rangle\leftrightarrow|1_mr_1\rangle\leftrightarrow|0_mr_2\rangle$ as shown in Fig.~2(c), can be described by the Hamiltonian
\begin{eqnarray}\label{e2}
	H_{11}={g}|0_mr_2\rangle\langle{1_mr_1}|+\frac{\Omega}{2}|1_mr_1\rangle\langle{1_m1_a}|+\rm H.c..
\end{eqnarray}
For a rectangular pulse and based on the time-evolution operator $U(t)=e^{-iH_{11}t}$, we can calculate the time-dependent probability amplitude of state $|1_m1_a\rangle$ as~\cite{SXF042}
\begin{eqnarray}\label{e3}
	C_{11}(t)=\frac{{4g}^2+{\Omega}^2 \cos(\frac{t}{2}\sqrt{4g^2+{\Omega^2}})}{{4g}^2+{\Omega}^2}.
\end{eqnarray}
\begin{figure}[htb]
	\centering
	\includegraphics[width=\linewidth]{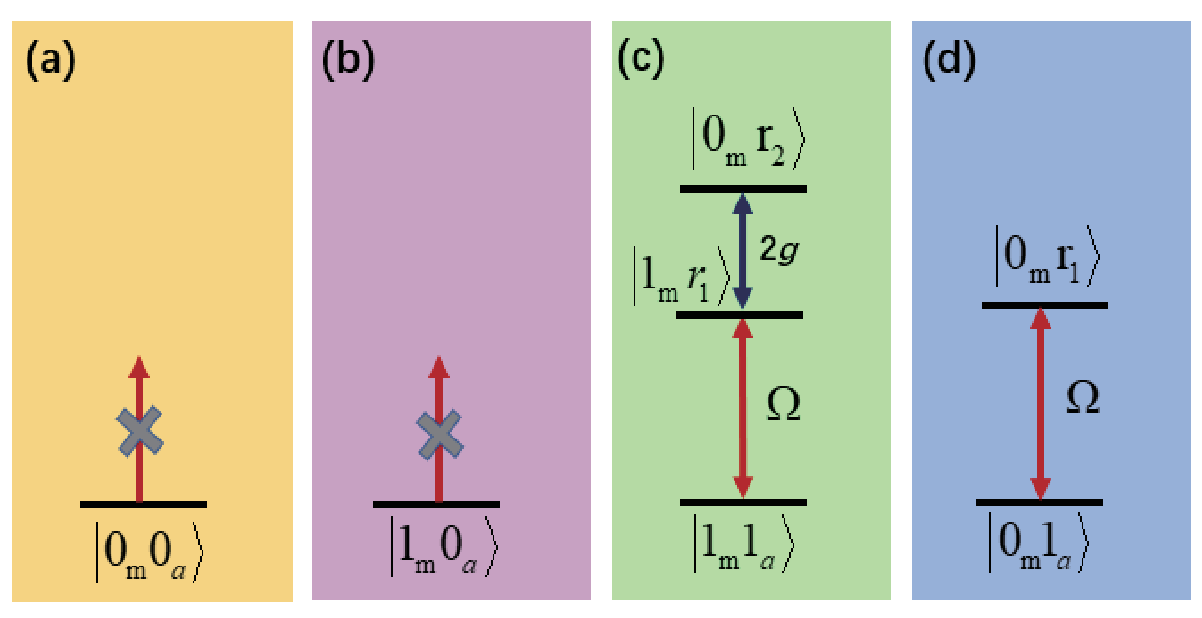}
	\caption{Gate processes, where $|0_m0_a\rangle$, $|0_m1_a\rangle$, $|1_m0_a\rangle$ and $|1_m1_a\rangle$ are the initial states of the atom-photon hybrid system.}
	\label{F1}
\end{figure}
If the system is initially in state $|0_m1_a\rangle$, the evolution of the hybrid system is governed by the Hamiltonian
\begin{eqnarray}\label{e4}
	H_{01}=\frac{\Omega}{2}\left|0_m1_a\right\rangle\left\langle0_mr_1\right|+\rm H.c.,
\end{eqnarray}
as depicted in Fig.~2(d). Given the time-independent Hamiltonian in Eq.~\eqref{e4}, the time-evolution operator is $U(t)=e^{-iH_{01}t}$, and the state $|0_m1_a\rangle$ undergoes Rabi oscillations as:
\begin{eqnarray}
	\left|\psi(t)\right\rangle=\cos(\frac{\Omega{t}}{2})\left|0_m1_a\right\rangle-i \sin(\frac{\Omega{t}}{2})\left|0_mr_1\right\rangle.\label{e5}
\end{eqnarray}

In the computational basis  \{$|0_m0_a\rangle$, $|0_m1_a\rangle$, $|1_m0_a\rangle$, $|1_m1_a\rangle$ \}, the CZ gate can be defined as the unitary transformation
\begin{eqnarray}\label{e6}
	U_{\mathrm{CZ}}=\left[\begin{array}{cccc}
		1 & 0 & 0 & 0 \\
		0 & e^{i \pi} & 0 & 0 \\
		0 & 0 & 1 & 0 \\
		0 & 0 & 0 & 1
	\end{array}\right].
\end{eqnarray}
Because the computational states $|0_m0_a\rangle$ and $|1_m0_a\rangle$ do not participate in the evolution, the CZ gate is realized when the state mappings $\left|1_m1_a\right\rangle\rightarrow\left|1_m1_a\right\rangle$ and $\left|0_m1_a\right\rangle\rightarrow-\left|0_m1_a\right\rangle$ are completed at the same time. Based on Eq.~\eqref{e3} and Eq. \eqref{e5}, we know that $\left|1_m1_a\right\rangle\rightarrow\left|1_m1_a\right\rangle$ and $\left|0_m1_a\right\rangle\rightarrow-\left|0_m1_a\right\rangle$ can be accomplished synchronously. For this, the gate time is set as $t=2\pi/\Omega$ while specific ratios $g/\Omega$ must be chosen, namely $g/\Omega=\sqrt{3}/2$, $g/\Omega=\sqrt{15}/2$, and $g/\Omega=\sqrt{35}/2\approx 2.9$ for the three smallest. As seen in Eq.~\eqref{e3}, in the limit $g/\Omega\gg1$, $C_{11}(t)\rightarrow 1$, such that in theory the effect of fluctuations on $g$ can be made arbitrarily low, and the gate operation is only sensitive to pulse duration via Eq.~\eqref{e5}. This result is due to the large dressing of the $|\phi_2 \rangle \leftrightarrow |\phi_3\rangle$ atomic transition by the strong microwave field that prevents any excitation in the Rydberg states. In practice, however, $g$ is limited by the geometry of the coplanar waveguide to a few MHz, and the ratio $g/\Omega$ cannot be made large enough~\cite{PAB010}. 

\section{CZ gate with Gaussian soft control}
Haase et al. introduced the GSC technique that
offers an effective tool for efficient RWA in a wide parameter regime~\cite{HJFWC050}. As shown in this section, it allows for efficient suppression of the unwanted excitation to the dressed states without the need for large $g/\Omega$ ratios. Following the GSC technique, the Rabi frequency starts from and ends at zero amplitude and maintains only on-resonant transitions. In practice, we use a varying Rabi frequency with Gaussian temporal profile of the form 
\begin{eqnarray}\label{e7}
	\Omega(t)=\Omega_me^{-(t-2\tau)^2/{\tau^2}},
\end{eqnarray}
to replace the time-independent $\Omega$. $\Omega_m$ and $\tau$ are the maximum amplitude and width of the Gaussian pulse, respectively. The time evolution of state $|0_m 1_a\rangle$ reads as in Eq. \eqref{e5}, where $\Omega t$ is replaced with the integral $\int_{0}^{t} \Omega (t') dt'$ for a given time $t$.
The gate duration is chosen as $T=4 \tau$ to nearly cover the entire Gaussian profile, and its pulse area verifies $\int_{0}^{T} \Omega (t') dt'\approx \sqrt{\pi} \Omega_m \tau$. 
From this, we can deduce that setting $\tau=2\sqrt{\pi}/\Omega_{m}$ is required for realizing the CZ gate.\\

To assess the performance of the CZ gate, we study the fidelity for preparing the atom-photon Bell state $\left|\rm {Bell}\right\rangle=\frac{1}{\sqrt{2}}(\left|0_m1_a\right\rangle+\left|1_m0_a\right\rangle)$  based on the numerical solution of the Hamiltonian of the hybrid system in Eq. \eqref{e1}.
For this, the system is assumed to be in the initial state
\begin{eqnarray}\label{e8}
	\left|\Psi_0\right\rangle=(\left|0_m\right\rangle+\left|1_m\right\rangle)/\sqrt{2}\otimes(\left|0_a\right\rangle+\left|1_a\right\rangle)/\sqrt{2}.\nonumber
\end{eqnarray}
Upon operating the CZ gate, $\left|\Psi_0\right\rangle$ should transform into the following state:
\begin{eqnarray}\label{e9}
	\left|\Psi_t\right\rangle=\frac{1}{\sqrt{2}}\left[\left|0_m\right\rangle \otimes\left(\left|0_a\right\rangle-\left|1_a\right\rangle\right)+\left|1_m\right\rangle\otimes\left(\left|0_a\right\rangle+\left|1_a\right\rangle\right)\right].\nonumber
\end{eqnarray}
Since $| \rm{Bell} \rangle= H_a | \Psi_t \rangle$ where $\rm{H_a}$ is the Hadamard gate applied to the atomic qubit, we can compute directly the fidelity of our gate as $F=\langle \Psi_t|\rho(t)| \Psi_t \rangle$, where $\rho(t)$ is the numerically computed density matrix of the system. This fidelity is equivalent to the fidelity of the Bell state preparation assuming that $\rm{H_a}$ is ideal.\\
\indent To choose a suitable ratio between g and $\Omega_m$ for achieving low error CZ gate with time-dependent $\Omega(t)$ based on GSC, we plot in Fig. 3 the fidelity of generating $| \rm{Bell} \rangle$ as a function of $g/\Omega_m$ at the time $T=4\tau=8\sqrt\pi/\Omega_m$. For comparison, the fidelity obtained at the time $t=2\pi/\Omega$ for the case of a time-independent rectangular pulse is also plotted in Fig. 3. 
\begin{figure}[htb]
	\centering
	\includegraphics[width=\linewidth]{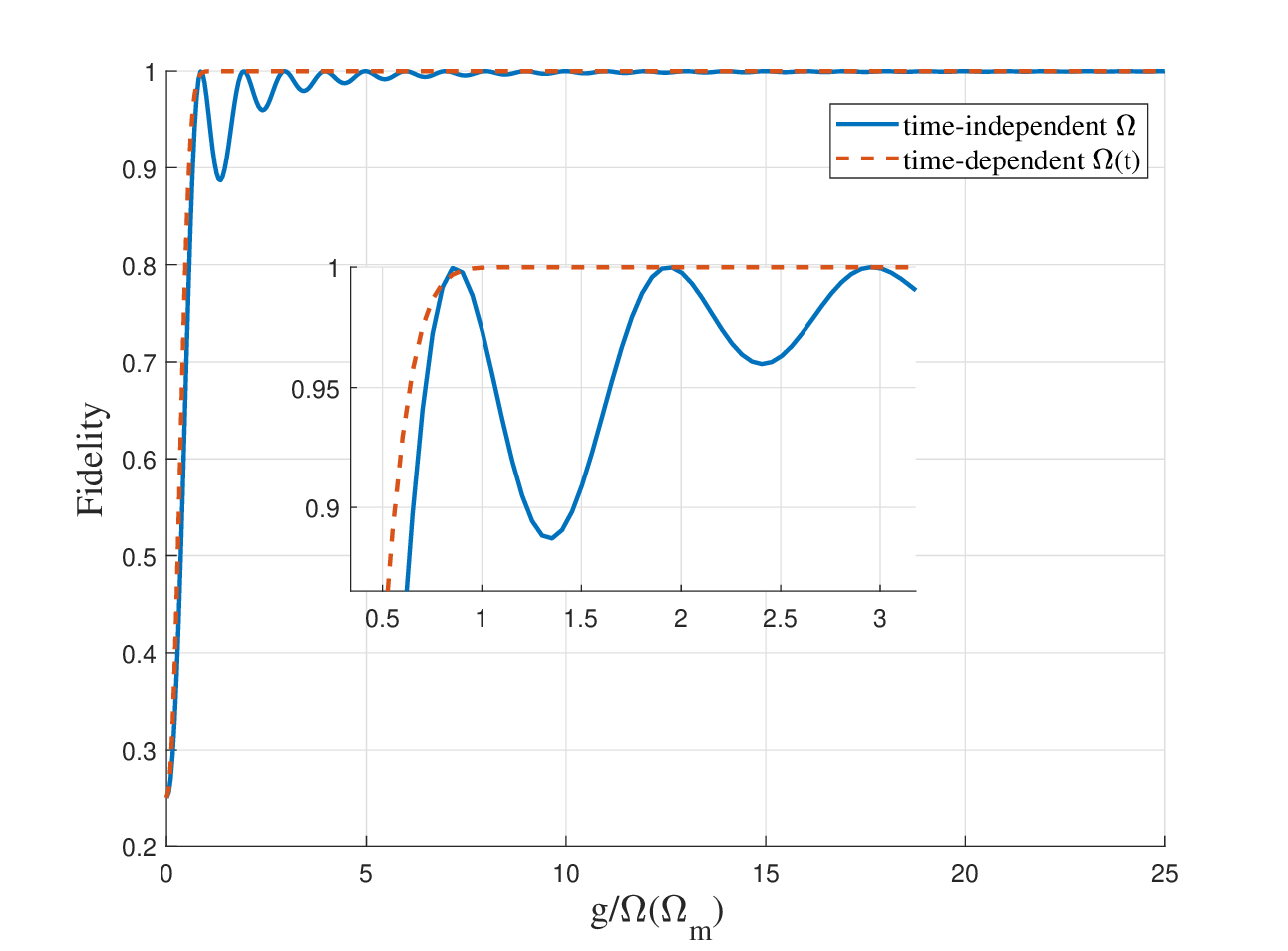}
	\caption{Influence of $g/\Omega(\Omega_m)$ on the gate fidelity. The gate fidelity is plotted versus $g/\Omega$ in the case of using a constant-amplitude pulse with gate time $t=2\pi/\Omega$ and versus $g/\Omega_m$ in the case of a time-dependent Gaussian pulse with gate time $T=8\sqrt{\pi}/\Omega_m$.}
	\label{F3}
\end{figure}
We observe that realizing a robust and high-fidelity $(F\textgreater0.99)$ CZ gate requires $g/\Omega_m\textgreater1$ when the GSC scheme is employed. Achieving a fast and high-fidelity CZ gate using a time-independent rectangular pulse requires choosing parameters on the maxima in Fig. 3, such as $g/\Omega=\frac{\sqrt3}{2}$, $\frac{\sqrt{15}}{2}$, and $\frac{\sqrt{35}}{2}$. With time-independent rectangular pulse, the fidelity can also keep larger than 0.99 in a wide range of $g$ when $g/\Omega\textgreater 8$ is fulfilled, however realizing this condition experimentally is very challenging, as explained in the previous section.\\
\indent Choosing $g/\Omega_m=1.3$ with time-dependent GSC modulation, we plot in Fig. 4 the time evolution of the populations and phases for $\left|1_m1_a\right\rangle$ and $\left|0_m1_a\right\rangle$ initial states. The populations are defined as $P_k=\left|C_k(t)\right|^2$, given the projection coefficient $C_k(t)$ along the computational state $|k\rangle$ $\left(k=1_m 1_a, 0_mr_2, 1_mr_1, 0_m1_a, 0_mr_1\right)$, and only the phase of $C_{k_0}(t)$ where $|k_0\rangle$ corresponds to the initial state is shown.
As expected, the populations in states $\left|1_m 1_a\right\rangle$ and $\left|0_m 1_a\right\rangle$ evolve back to their initial values. The phase $\arg(C_{1_m 1_a})$ corresponding to the $\left|1_m 1_a\right\rangle$ initial state is unmodified, whereas the real-valued coefficient $C_{0_m 1_a}$ related to the initial state $\left|0_m 1_a\right\rangle$ acquires a stable phase shift of $\pi$ when it changes sign, which ensures that the CZ gate is robust against large fluctuations of the gate time $T$. Furthermore, using the time-dependent Gaussian pulse suppresses the population in the Rydberg state $|r_1\rangle$ when $\left|1_m1_a\right\rangle$ is the initial two-qubit state, such that the effect of its finite lifetime on the fidelity may be reduced.\\
\begin{figure}[htb]
	\centering
	\includegraphics[width=\linewidth]{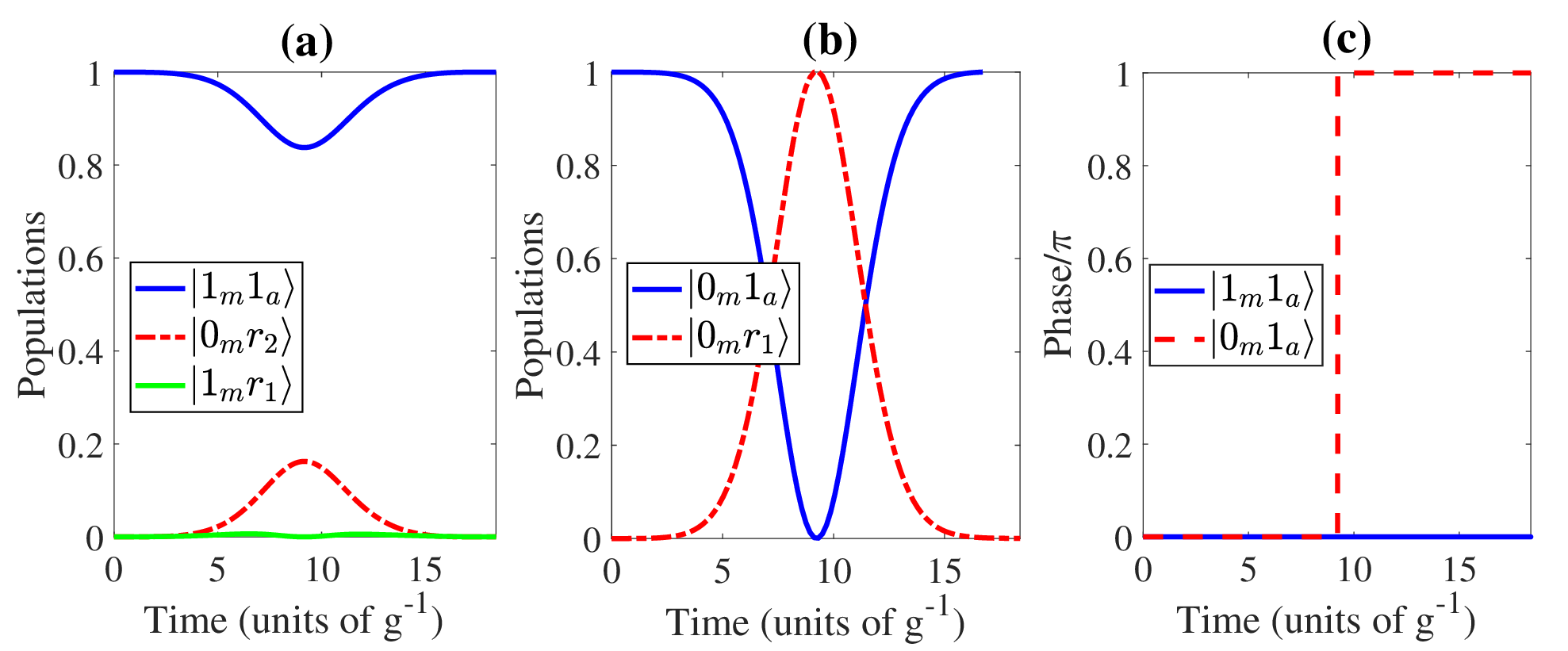}
	\caption{Time evolutions of population and phase when the Rabi frequency is driven by the GSC pulse with $g/\Omega_m=1.3$. (a) Populations of states $\left|1_m1_a\right\rangle$, $\left|0_mr_2\right\rangle$, $\left|1_mr_1\right\rangle$ when $\left|1_m1_a\right\rangle$ is the initial state. (b)  Populations of states $\left|0_m1_a\right\rangle$ and $\left|0_mr_1\right\rangle$ when $\left|0_m1_a\right\rangle$ is the initial state. (c) Phase evolutions of $\left|0_m1_a\right\rangle$ and $\left|1_m1_a\right\rangle$ when $\left|0_m1_a\right\rangle$ and $\left|1_m1_a\right\rangle$ are the initial states, respectively.}
	\label{F4}
\end{figure}  
The shortest gate with time-independent pulse is obtained when $g/\Omega=\frac{\sqrt3}{2}$. However, to further characterize the GSC method, Fig. 5 plots the time evolution of the fidelity with $g/\Omega_m=1.3$ for the GSC pulse and $g/\Omega=2.9 (\approx\sqrt{35}/2)$ for the time-independent rectangular pulse to ensure using almost the same gate time of $T=18.4/g$. 
As shown in Fig. 5, the CZ transformation with GSC occurs during a reduced time window as compared to the case of the time-independent rectangular pulse, which reduces the sensitivity of the gate to dissipation and laser phase noise. Furthermore, the fidelity is stable at the end of the Gaussian pulse.
\begin{figure}[!h]
	\centering
	\includegraphics[width=\linewidth]{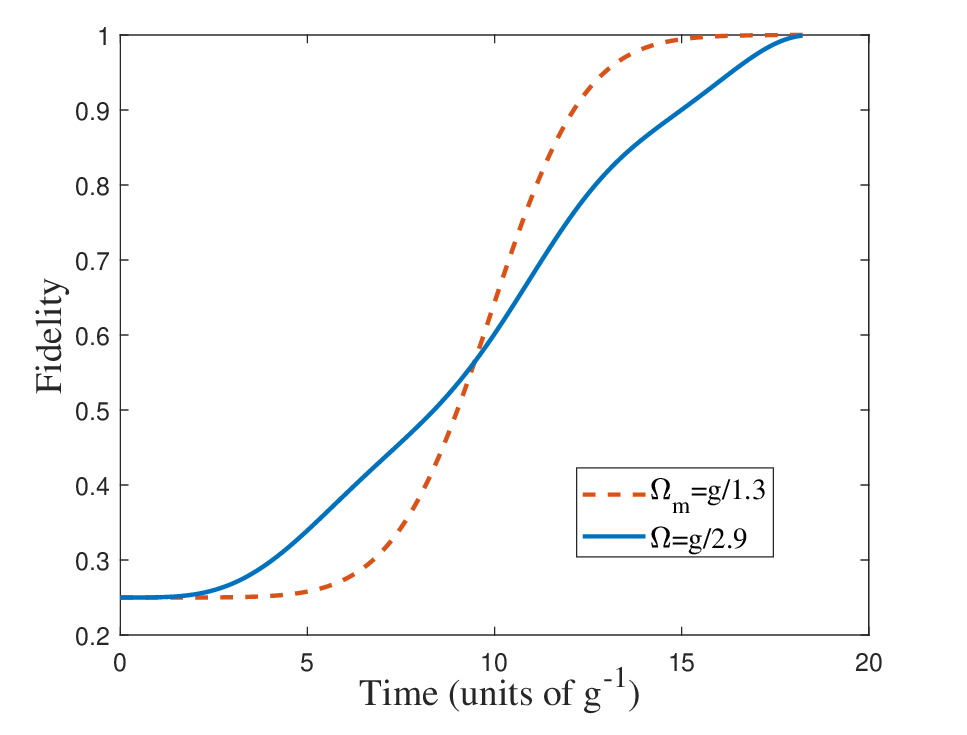}
	\caption{Time evolutions of the gate fidelity for the case of GSC pulse with $\Omega_m=g/1.3$ and time-independent pulse with $\Omega=g/2.9$.}
	\label{F5}
\end{figure}
\section{Robustness against parameter imprecision and decoherence}
As mentioned in the previous section, compared with the gate scheme based on a  rectangular pulse, the GSC one is resistant to deviations of the gate time $T$ and atom-photon coupling strength $g$. To further illustrate these characteristics, we investigate the effect of variations of $T$ and $g$ on the Bell state fidelity. For a given parameter $X \in \{g,T\}$, its relative error is defined as $\delta{X}/X$, $\delta{X}$ being the deviation value from $X$. Figs. 6(a) and 6(b) simulate the effect of the relative errors in $T$ and $g$ on the Bell state fidelity in both cases of time-independent and GSC pulses. This simulation is performed with the parameters of Fig.~5, and we also set $g/\Omega=\frac{\sqrt3}{2}$ to make a comparison with Ref. \cite{SXF042}.
\begin{figure}[htb]
	\centering
	\includegraphics[width=\linewidth]{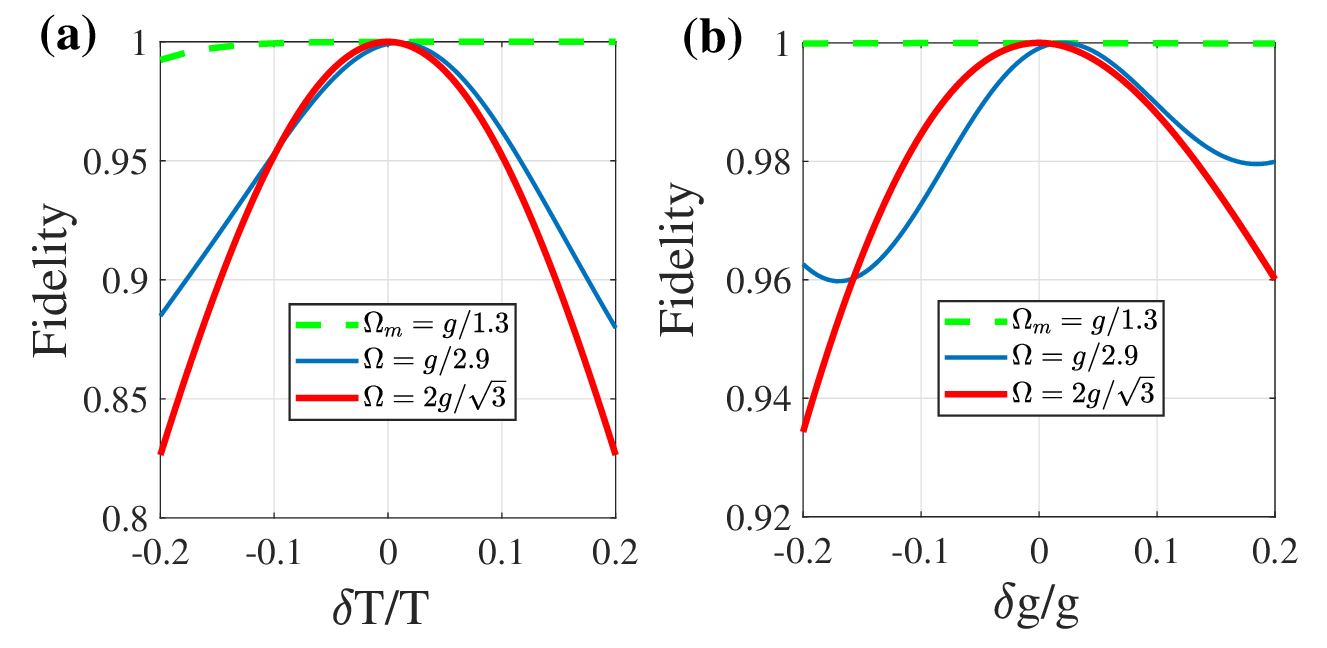}
	\caption{Influence on the gate fidelity of relative errors in (a) gate time and (b) atom-cavity field coupling strength.}
	\label{F6}
\end{figure}
In Fig. 6, we observe that the gate scheme that relies on a rectangular pulse with $g/\Omega=\frac{\sqrt3}{2}$ is significantly influenced by fluctuations of $g$ and $T$. On the contrary, the Bell state fidelity obtained with GSC is insensitive to both $g$ and gate duration errors.\\
\indent One of the main issues related to SCW resonators is that large stray electric fields occur near the surface of the SCW, of up to tens of V/cm at a distance of 10 \textmu m if alkali atoms are chosen~\cite{KMGL013}. These stray electric fields $F$ induce large Stark shifts of the Rydberg energy levels, typically of the form $-\alpha F^2/2$ for a state of low orbital angular momentum $\ell$, where the polarizability $\alpha$ depends on $n$ and $\ell$ ~\cite{GTFR2000}. They can be compensated at a given position above the SCW, however quite imperfectly, such that large inhomogeneities of the detuning remain within the atom trap.  Hence, our gate scheme should consider this issue carefully.
Fig. 7 displays the fidelity of the gate as a function of the detuning error of the two Rydberg energy levels $|r_1\rangle$ and $|r_2\rangle$ induced by the presence of stray fields. As we can see from Fig. 7, considering that $|r_1\rangle$ and $|r_2\rangle$ are $nS$ and $nP$ Rydberg states of cesium with polarizabilities of $\alpha_1$ and $ \alpha_2\approx \alpha_1 \times 22$, respectively, the gate scheme based on the GSC technique is very robust against the Stark-shift-induced detunings $\Delta_2= \alpha_2 F^2/2$ of the microwave field. Note that the detuning $\Delta_1=\alpha_1/\alpha_2 \Delta_2 $ is taken into account in the simulation for consistency.
This central result is expected as the Raman transition  $|1_m1_a\rangle\leftrightarrow|1_mr_1\rangle\leftrightarrow|0_mr_2\rangle$ is nearly suppressed when $\Omega_m=g/1.3$ is satisfied, hence the gate depends little on the presence of a detuning of the Rydberg state $|r_2\rangle $, and only detuning of state $|r_1\rangle$ plays an important role via the two-level Rabi oscillations of Eq.~\eqref{e5}. A state $|r_1\rangle$ with large polarizability such as $nP$ can be chosen, however a reduced polarizability state is still necessary for $|r_1\rangle$. As shown in Fig. 7, detunings for the $|r_2\rangle$ state of up to $0.5g$ are acceptable, that is of up to $~2\pi\times 1$~MHz. \\
\begin{figure}[htb]
	\centering
	\includegraphics[width=\linewidth]{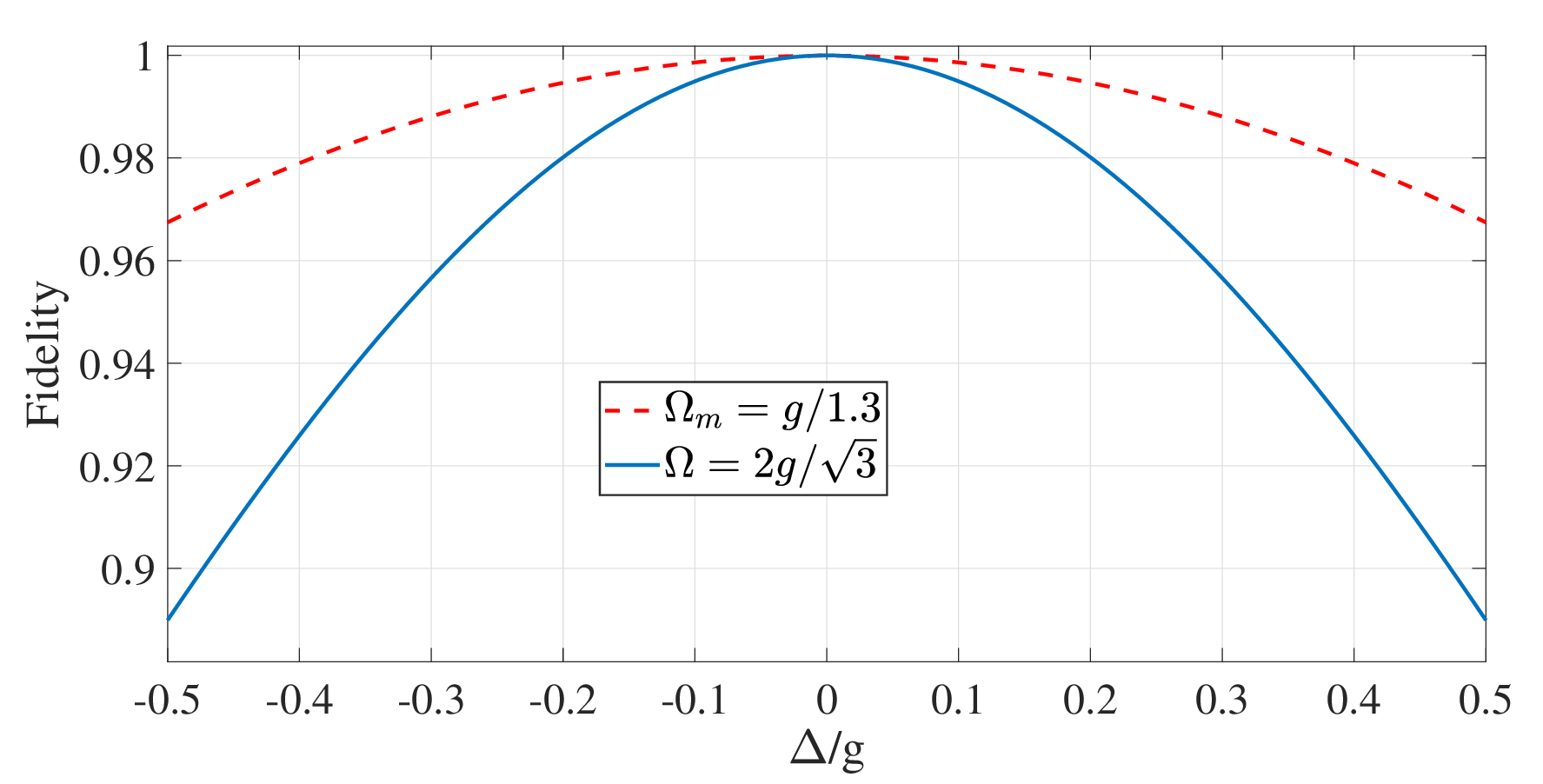}
	\caption{Influence of the detuning error $\Delta=\Delta_2$ of the Rydberg state $|r_2\rangle$ on the gate fidelity. Detuning of state $|r_1\rangle$ is considered to be proportional to $\Delta$  (see text).}
		\label{F7}
	\end{figure}
\indent Furthermore, we consider the effect of decoherence caused by dissipation in the hybrid system. The relaxation processes of our hybrid atom-photon system are inevitable. Here we take into account two decoherence factors, including the spontaneous emission of the atom and the cavity photon decay. The spontaneous decays can be represented with a Lindblad super operator acting on the density operator $\hat{\rho}$ of the hybrid system as 
\begin{eqnarray}\label{e10}
	\mathcal{L}_a\hat{\rho}=\frac{\gamma_1}{2}\left(2\left|1_a\right\rangle\left\langle r_1|\hat{\rho}| r_1\right\rangle\left\langle 1_a|-| r_1\right\rangle\left\langle r_1|\hat{\rho}-\hat{\rho}| r_1\right\rangle\left\langle r_1\right|\right)\nonumber
\end{eqnarray}
\begin{eqnarray}\label{e10b}
	+\frac{\gamma_2}{2}\left(2\left|1_a\right\rangle\left\langle r_2|\hat{\rho}| r_2\right\rangle\left\langle 1_a|-| r_2\right\rangle\left\langle r_2|\hat{\rho}-\hat{\rho}| r_2\right\rangle\left\langle r_2\right|\right),
\end{eqnarray}
where $\gamma_{1(2)}$ is the decay rate of the Rydberg state $|r_{1(2)}\rangle$. For convenience, we assume that $\gamma_1=\gamma_2=\gamma$.
The cavity photon decay process can be expressed as 
\begin{eqnarray}\label{e11}
	\mathcal{L}_c \hat{\rho}=\frac{\kappa}{2}(2 \hat{a} \hat{\rho} \hat{a}^{\dagger}-\hat{a}^{\dagger} \hat{a} \hat{\rho}-\hat{\rho} \hat{a}^{\dagger} \hat{a}),
\end{eqnarray}
where $\kappa$ is the decay rate of the microwave photon in the SCW cavity and $\hat{a}$ corresponds to the
annihilation operator of the SCW microwave field.\\
The dynamics of the dissipative system is described by the following Markovian master equation 
\begin{eqnarray}\label{e12}
	\dot{\rho}=-\frac{i}{\hbar}[H_I, \rho]+\mathcal{L}_a \hat{\rho}+\mathcal{L}_c \hat{\rho}.
\end{eqnarray}
\begin{figure}[htb]
	\centering
	\includegraphics[width=\linewidth]{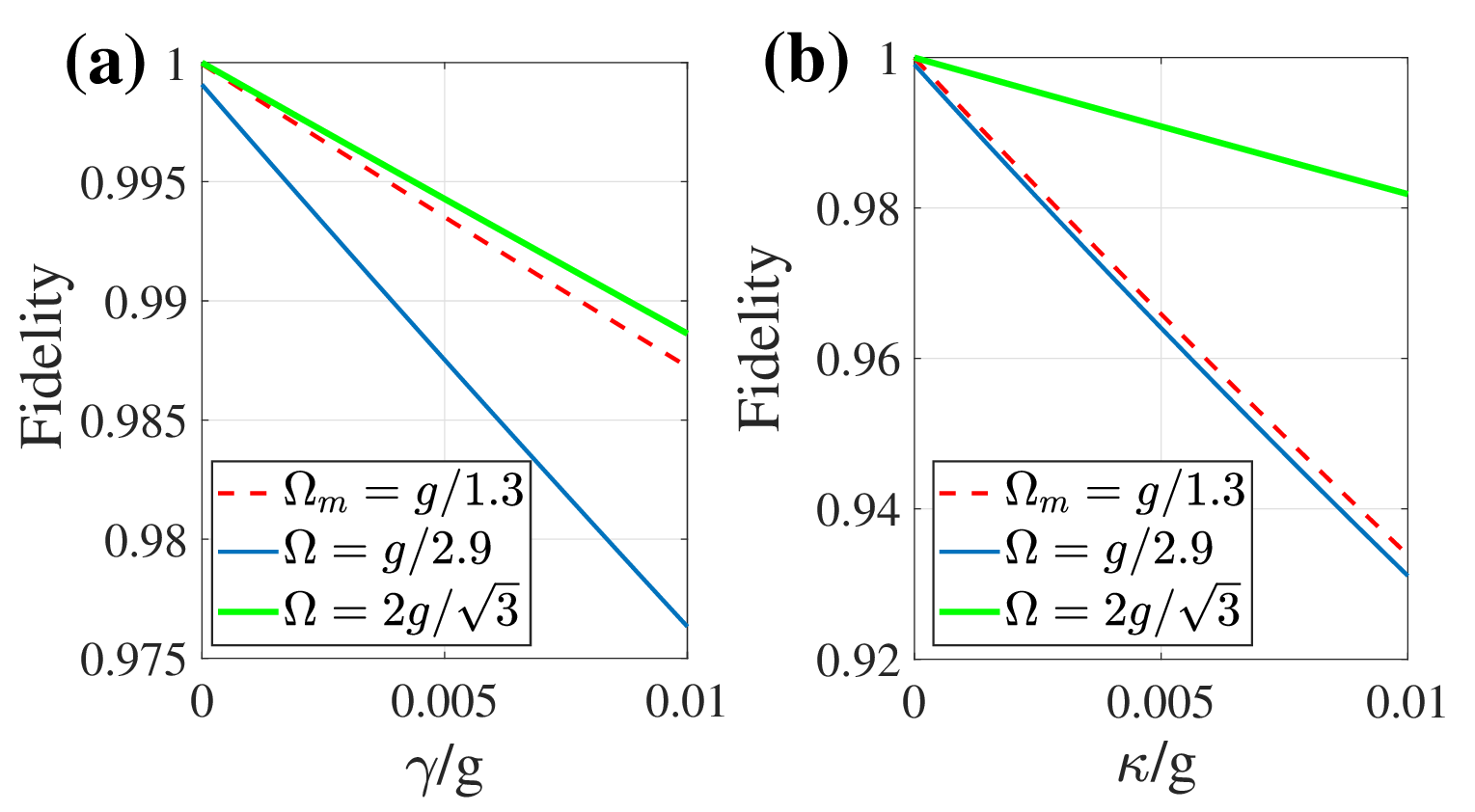}
	\caption{Influences of decay errors on the fidelity for both cases of rectangular and GSC pulses: (a) Effect of spontaneous emission rate $\gamma$; (b) Effect of the cavity decay rate $\kappa$.}
	\label{F8}
\end{figure}
	From the numerical solution of Eq.~\eqref{e12}, the dependence of the Bell state fidelity is plotted in Fig.~8 as a function of the decoherence rates $\gamma$ and $\kappa$ for time-dependent Gaussian and time-independent rectangular pulses. In both cases, we observe that $\kappa$ has a stronger influence on the Bell-state fidelity than $\gamma$. When considering the evolutions of the fidelity versus $\gamma$ and $\kappa$ in Fig.~8 at fixed gate time ($T=18.4/g$ for $\Omega_m=g/1.3$ and $\Omega=g/2.9$), we see that the GSC scheme is more robust against the decays than the rectangular pulse scheme. Nevertheless, such is not case for the fastest gate based on the rectangular pulse that relies on the ratio $\Omega=2g/\sqrt{3}$.\\ 
\indent Greater ratios of $g/\Omega(\Omega_m)$ conduce to longer gate times, larger decoherence, and smaller fidelity.
However, not only the gate time matters but also the probability of excitation into the Rydberg states.  Although the gate time corresponding to $\Omega_m=g/1.3$ in the case of GSC is nearly 3.3 times longer than that corresponding to $\Omega=2g/\sqrt{3}$ for a rectangular pulse, it shows almost the same resilience to spontaneous emission because of the relatively small populations in the excited states that was reported in Fig. 4(a).
The main drawback of the GSC technique is its stronger sensitivity to cavity photon decay that is due to the required longer gate time. The photon decay rate is related to the quality factor of the cavity $Q$ and its frequency $\omega_c $ as $\kappa =\omega_c/Q$,  hence our scheme is mostly useful for quality factors $Q\geq  10^6$ in the range of $\omega_c$ typically used in SC systems. Such high quality factors have not been reported yet at temperature $\textless$1 K but this target seems reasonable, as much improvement has been realized recently~\cite{MANC2012}.\\
\indent As a complete example, we compute the fidelity based on the setup analyzed in Ref.~\cite{SXF042}. Two Rydberg states are chosen as the $\left|\mathrm{r}_1\right\rangle \equiv\left|90 S_{1 / 2,} m_J=1 / 2\right\rangle$ and $\left|\mathrm{r}_2\right\rangle \equiv\left|90 P_{3 / 2,} m_J=1 / 2\right\rangle$ states of cesium. When the temperature is below 400~mK, the lifetime of the two Rydberg states are $\tau_1=0.82\ ms$ and $\tau_2=1.97~ms$, and the corresponding decay rates are $\gamma_1/2\pi=194\ \rm{Hz}$ and $\gamma_2/2\pi=80\ \rm{Hz}$, respectively. The SCW frequency $\omega_c=2\pi\times5.037\ \rm{GHZ}$ resonantly couples the transition between $\left|\mathrm{r}_1\right\rangle$ and
$\left|\mathrm{r}_2\right\rangle$. Assuming that the CPW resonator has a quality factor of $Q=10^6$, the photon decay rate $\kappa$ is given by $\kappa=\omega_c/Q\simeq2\pi\times5$ ${\rm{kHz}}$. Trapping of atoms at a height of 10~$\mu m$ above the SCW resonator achieves a coupling strength of $g/2\pi\simeq2\ \rm{MHz}$ for $n=90 $\cite{PAB010}. Considering the conditions discussed previously, we set $\Omega_m=g/1.3\simeq1.54\times2\pi$ ${\rm{MHz}}$. Based on these parameters and on the simulation of Fig. 7, we find that the fidelity exceeds 0.98 and can be further improved with a greater quality factor.
\section{Conclusion}
In summary, we have proposed a theoretical scheme to achieve a resilient atom-photon CZ gate in a SCW resonator-atom hybrid quantum system. The gate protocol to be implemented is simple because it employs only one classical field to induce a transition between the ground and Rydberg states for constructing the CZ gate. Introducing Gaussian temporal modulation to maintain on-resonant transitions, the scheme demonstrates a high level of robustness against control errors in gate time, atom-photon coupling strength, and stray electric fields. The gate based on GSC is resilient against spontaneous decay rates, but is more sensitive to cavity photon decay compared to the gate relying on a rectangular pulse. For avoiding this latter problem, a cavity with $Q\textgreater10^6$ will have to be employed. The hybrid atom-photon CZ gate can be used to generate entangled states between atoms and microwave photons. We believe that this approach holds great promise for future quantum networks. 

\section*{Acknowledgment}
This work is supported by the National Natural Science Foundation of China under grant number 12174460.


\end{document}